%% file: KacPoly.tex
\documentclass[a4paper,12pt]{article}

\usepackage{amsmath, amsthm, amssymb}
\usepackage{graphicx}

\begin{document}
\title{The limiting Kac random polynomial and truncated random orthogonal matrices}
\author{Peter J. Forrester}
\date{}
\maketitle

\noindent
\thanks{\small
Department of Mathematics and Statistics,
The University of Melbourne, 
Victoria 3010, Australia \\[.1cm]
email: { P.Forrester@ms.unimelb.edu.au}

\begin{abstract}
\noindent
An exact  calculation of  the eigenvalue statistics of truncated random Haar distributed real orthogonal matrices has recently been carried out by Khoruzhenko, Sommers and 
Zyczkowski. We further develop this calculation, and use it to deduce  a Pfaffian form of the correlations for the zeros of the limiting Kac random polynomial. This contrasts with the forms known from previous studies of the real zeros (a multidimensional Gaussian integral with the integrand multiplied by the absolute values of the variables) and the complex zeros (a Hafnian).
\end{abstract}

\section{Introduction}
There has been much recent progress in the exact evaluation of correlation functions for various non-Hermitian, real entried random matrix ensembles \cite{FN07,FN08p,Som07,SW08,FM08,
APS10,AKP10,KSZ10}. The most recent of these works, due to Khoruzhenko et al.~\cite{KSZ10},
lays the foundations to completely solve this problem for the ensemble of real entried random matrices formed from
$M \times M$ sub-blocks of $(L+M) \times (L+M)$ Haar distributed real orthogonal matrices. In light of the recent work of Krishnapur \cite{Kr06}, the case $L=1$, $M \to \infty$, has immediate implication to the evaluation of the $N \to \infty$ limiting correlation functions for the zeros of the so-called Kac random polynomial
\begin{equation}\label{0}
f(x) = \sum_{p=0}^N a_p x^p,
\end{equation}
where each $a_p$ is an independent standard real Gaussian. It is the purpose of this paper to make these implications explicit, and to similarly make explicit the implications for the point process
defined by the zeros of the matrix-valued real Gaussian power series
\begin{equation}\label{1}
\det \Big ( \sum_{p=0}^\infty A_L^{(p)} x^p \Big ) = 0.
\end{equation}
Here each $A_L^{(p)}$ is an independent $L \times L$ standard Gaussian real random matrix.

We begin in Section 2 by adapting the work of Krishnapur \cite{Kr06} to show the relationship between the eigenvalues of truncations of random orthogonal matrices, and the zeros of 
(\ref{1}). We further revise known exact expressions, due to Bleher and Di \cite{BD97}, for the
correlations of the real zeros of (\ref{0}), and known exact expressions due to Prosen \cite{Pr96}
for the correlations of the complex zeros. In Section 3 we extend the workings in \cite{KSZ10}
in relation to the exact evaluation of the correlations for eigenvalues of truncations of random
orthogonal matrices. Thus we give the explicit $M \to \infty$ form of the entries of the Pfaffian specifying this correlation function.

The result from Section 2 giving an equivalence between the eigenvalues of certain truncated
orthogonal matrices and the limiting zeros of (\ref{1}), together with the Pfaffian form of the eigenvalue correlations obtained in Section 3 tell us that the limiting zeros of the Kac polynomial form a Pfaffian point process. However, neither the correlations between the real zeros, nor the correlations between the complex zeros obtained in \cite{BD97} and \cite{Pr96} respectively, are in a Pfaffian form.
We can check directly that the one and two point function of the former, and the one point function of the latter, coincide. This is the essence of our conclusion, Section 4, which therefore leaves open the problem of directly showing the equivalence between the different forms.

\section{Preliminary results}
\subsection{The result of Krishnapur}
Consider the ensemble $\{U\}$ of Haar distributed $(L+M) \times (L+M)$
real orthogonal matrices. Block decompose $U$
according to
\begin{equation}\label{UA}
U = \begin{bmatrix} A_{L \times L} & B \\
C & V_{M \times M} \end{bmatrix}.
\end{equation}
As noted in \cite{Kr06} in the case of complex unitary matrices, these blocks satisfy the determinant
identity
\begin{equation}\label{3}
{\det(z \mathbb I_M - V^T) \over \det (\mathbb I_M - z V) } =
(-1)^M \det U
\det ( A + z B (\mathbb I_M - z V)^{-1} C).
\end{equation}

Now write
\begin{equation}\label{3a}
f_M(z) = (-1)^M \det U {\det (z \mathbb I_M - V) \over \det (\mathbb I_M - z V) }.
\end{equation}
According to (\ref{3}),
\begin{eqnarray}\label{3b}
M^{L/2} f_M(z) &=& \det \Big ( \sqrt{M} (A + zB (\mathbb I_M - z V)^{-1} C) \Big ) \nonumber \\
&& \times \det \Big (\sqrt{M} ( A + z B C + z^2 B V C + z^3 B V^2 C + \cdots) \Big ).
\end{eqnarray}
But we know from \cite[pg.~113]{HKPV08} that
\begin{equation}\label{3c}
\sqrt{M}(A,BC,BVC,BV^2C,\dots ) \mathop{\sim}\limits^{\rm d}
(A_L^{(0)},A_L^{(1)},A_L^{(2)},\dots),
\end{equation}
and furthermore that for $M \to \infty$ this implies the eigenvalues of the sub-block $V$ coincide with the zeros of (\ref{1}) for $N \to \infty$ (compare (\ref{3a}) with (\ref{3b}), (\ref{3c})).

\subsection{Limiting real and complex correlations for the Kac polynomial}
In relation to (\ref{0}), it was proved by Kac \cite{Ka43} that for large $N$ the leading
order expected number of real zeros is $(2/\pi) \log N$. Also, it is known
(see \cite{BD97} and references therein) that for $N \to \infty$ the density of real eigenvalues is
\begin{equation}\label{4}
\rho_{(1)}^{\rm r}(x) = {1 \over \pi |1 - x^2|}.
\end{equation}
This is singular at $x = \pm 1$ --- the points where the zeros accumulate, but is otherwise well defined.

An important point from \cite{BD97} is that the zeros for $|x| < 1$ are statistically independent from the zeros for $|x| > 1$. This is in keeping with the fact \cite[Lemma 2.3.3]{HKPV08} that for $N \to \infty$ the random analytic function (\ref{1}) has radius of convergence unity, and moreover cannot be analytically continued to $|x|\ge1$.. It also explains why it is consistent that the eigenvalues of $V$ in Section 2.1 all have modulus less than 1.

The general $k$-point correlation function $\rho_{(k)}^{\rm r}$ for the real zeros in $(-1,1)$
of the limiting Kac
polynomial (\ref{0}) has been computed by Bleher and Di \cite{BD97}. Furthermore, these authors have shown that this same functional form persists for a wide class of mean zero, variance one distributions for the coefficients $a_p$ in (\ref{0}) \cite{BD04}. To state the result,
let
$$
\Delta_k = [ \Delta_{jl} ]_{j,l=1,\dots,k}
$$
with each $\Delta_{jl}$ a $2 \times 2$ block
$$
\Delta_{jl} = \begin{bmatrix} \displaystyle{1 \over 1 - x_j x_l} & 
\displaystyle{x_j \over (1 - x_j x_l)^2} \\
\displaystyle {x_l \over (1 - x_j x_l)^2} & \displaystyle {1 + x_j x_l \over (1 - x_j x_l)^3} \end{bmatrix}.
$$
Then we have from \cite[eq.~(4.6)]{BD97} that
\begin{eqnarray}\label{rr}
&& \rho_{(k)}^{\rm r}(x_1,\dots,x_k) \nonumber \\
&& \qquad = {1 \over (2 \pi)^k \sqrt{\det \Delta_k}}
\int_{-\infty}^\infty dy_1 \cdots \int_{-\infty}^\infty dy_k \,
|y_1 \cdots y_k| e^{ - \vec{y}^T \Omega_k \vec{y}},
\end{eqnarray}
where $\vec{y} = [y_l]_{l=1,\dots,k}$ is a column vector and $\Omega_k$ denotes the $k \times k$ matrix obtained by removing all the odd number rows and columns from $\Delta_k^{-1}$.
In the case $k=1$ this reclaims (\ref{4}), while for $k=2$ it simplifies to give
\cite[eq.~(3.10)]{BD97}
\begin{eqnarray}\label{4a}
\rho_{(2)}^{\rm r}(x_1,x_2) & = & {(x_1 - x_2)^2 \over \pi^2 (1 - x_1 x_2)^2 (1 - x_1^2)(1-x_2^2)} \nonumber
\\
&&\! \! + {|x_1 - x_2| \over \pi^2 (1 - x_1 x_2)^2 \sqrt{(1 - x_1^2)(1 - x_2^2)}}
\arcsin {\sqrt{(1 - x_1^2)(1 - x_2^2)} \over 1 - x_1 x_2}
\end{eqnarray}
(for an extended discussion of issues relating to a direct evaluation of the integral
in (\ref{rr}) for general $\Omega_k$, see \cite{LW09}).

We now turn our attention to the complex zeros. For the $a_p$ in (\ref{0}) any independent real Gaussian distribution
with variance $b_p$, and thus
\begin{equation}\label{ab}
\langle a_p a_q \rangle = b_p \delta_{p,q},
\end{equation}
a formula for the $k$-point correlation $\rho_{(k)}^{\rm c}$ of the complex zeros has been given by Prosen \cite{Pr96}. Since the coefficients of the polynomials are real, the complex zeros must occur in complex conjugate pairs. The exact evaluation is given in terms of a quantity known as a Hafnian
(see e.g.~\cite{IKO05}) but referred to as a semi-permanent in \cite{Pr96},
\begin{equation}\label{HA}
{\rm Hf} \, A_{2k \times 2k} = \sum_{P \in S_{2k}} \! \!{}^* \, \prod_{l=1}^k
a_{P(2l-1), P(2l)}.
\end{equation}
Here $A_{2k \times 2k}  = [a_{jk}]_{j,k=1,\dots,2k}$ is assumed symmetric, and the asterisk on the sum means that the permutations must statisfy $P(1) < P(3) < \cdots < P(2k-1)$ and
$P(2i-1) < P(2i)$ ($i=1,\dots,k$). According to \cite[eq.~(12)]{Pr96} we have that
\begin{equation}\label{Hf}
\rho_{(k)}^{\rm c}(z_1,\dots,z_k) = {{\rm Hf} \, (C - B^\dagger A^{-1} B) \over
\sqrt{\det (2 \pi A)}}.
\end{equation}
The matrices $A,B,C$ are of size $2k \times 2k$ and have entries defined in terms of
\begin{equation}\label{gu}
g(u) := \sum_{n=0}^Nb_n u^n,
\end{equation}
where $b_n$ is specified by (\ref{ab}). Explicitly, with $z_{k+s} := \bar{z}_s$
and $g'(u)$ denoting the derivative of $g(u)$,
these matrices have entries
\begin{align*}
& A_{jl} = g(z_j \bar{z}_l) \\
& B_{jl} = \partial_{\bar{z}_l} A_{jl} = z_j g'(z_j \bar{z}_l) \\
& C_{jl} = \partial_{z_j} \partial_{\bar{z}_l} A_{jl} = g'(z_j \bar{z}_l) +
z_j \bar{z}_l g''(z_j \bar{z}_l).
\end{align*}
In the present problem we have $b_n = 1$ for each $n$, and we must also take
the limit $N \to \infty$. Then (\ref{gu}) simplifies to read
\begin{equation}\label{g}
g(u) = {1 \over 1 - u}.
\end{equation}

In the case $k=1$ (\ref{Hf}) can be written out explicitly (\cite[eq.~(19)]{Pr96} for general $g$.
Substituting (\ref{g}) in that formula we find
\begin{equation}\label{zc}
\rho_{(1)}^{\rm c}(z) = {1 \over \pi} {|z - \bar{z}| \over |1 - z^2| (1 - |z|^2)^2}.
\end{equation}

\section{Pfaffian correlations for eigenvalues of truncated orthogonal matrices}
\subsection{Specification of the correlations}
Let an $(L+M) \times (L+M)$ random Haar distributed real orthogonal matrix $U$ be decomposed as in (\ref{UA}). The exact form of the joint eigenvalue PDF and the $k$-point correlations has been the subject of the recent work \cite{KSZ10}. It was shown that up to a known normalization, the joint eigenvalue PDF is given by
\begin{equation}\label{zc1}
\prod_{1 \le j < k \le M} (z_j - z_k) \prod_{j=1}^M w(z_j)
\end{equation}
where
\begin{equation}\label{w}
w(z) = \left \{\begin{array}{ll} \displaystyle \Big ( {L(L-1) \over 2 \pi} |1 - z^2|^{L-2}
\int_{2|y|/|1 - z^2|}^1 (1 - u^2)^{(L-3)/2} \, du \Big )^{1/2}, & L \ne 1 \\
\displaystyle\Big ( {L \over 2\pi} \Big )^{1/2} |1 - z^2|^{-1/2},& L = 1. \end{array} \right.
\end{equation}
Note that when $z$ is real and thus $y=0$, (\ref{w}) redues to
\begin{equation}\label{wr}
w(x) = \Big ( {L \over 2 \pi} {\Gamma(1/2) \Gamma((L+1)/2) \over \Gamma(L/2)} \Big )^{1/2}
(1 - x^2)^{L/2 - 1}.
\end{equation}

The functional form (\ref{zc1}) must be considered as a sum over all allowed numbers of real
eigenvalues. Thus with $M$ even (for convenience) there can be an even number
of real eigenvalues, while the remaining eigenvalues come in complex conjugate pairs.
As noted in \cite{KSZ10}, but not made explicit, the structured form (\ref{zc1}) implies that the correlation functions between general sets of $k_1$ real eigenvalues and $k_2$ complex eigenvalues (the latter with imaginary part positive) can be written as a
$2(k_1 + k_2) \times 2(k_1 + k_2)$ Pfaffian \cite{FN07,BS07,FM08,}.

The key to such formulas is an antisymmetric inner product induced by (\ref{zc1}),
\begin{eqnarray}\label{g1g}
\langle g_1, g_2 \rangle & := & 2i \int \int_{D^+} (w(z))^2 \Big ( g_1(z) g_2(\bar{z}) -
g_1(\bar{z}) g_2(z) \Big ) dx dy \nonumber \\
&& + \int_{-1}^1  \int_{-1}^1  w(x_1) w(x_2) g(x_1) g(x_2) {\rm sgn} \, (x_2-x_1) \, dx_1 dx_2
\end{eqnarray}
where $z = x + i y$ and $D^+$ is the upper half unit disk.

Let $\{p_j(z) \}_{j=0,1,\dots}$ be a set of skew orthogonal polynomials with respect to (\ref{g1g}),
\begin{eqnarray}\label{sk}
&&\langle p_{2j},p_{2k} \rangle = \langle p_{2j+1},p_{2k+1} \rangle = 0 \nonumber \\
&& \langle p_{2j},p_{2k+1} \rangle = 0 \quad (j \ne k), \qquad  \langle p_{2j},p_{2j+1} \rangle = r_j,
\end{eqnarray}
where it is required that each $p_j(z)$ is monic and of degree $j$. Use these monic polynomials and the weight (\ref{w}) to define
\begin{eqnarray}
\tilde{w}(u) & := & \sqrt{2} w(u) \\
q_j(u) & := & \tilde{w}(u) p_j(u) \label{q1}\\
\tau_j(u) & = & \left \{ \begin{array}{ll}\displaystyle -{1 \over 2} \int_{-1}^1 {\rm sgn} \,(u-t) q_j(t) \, dt,
& u \in \mathbb R \\
i q_j(\bar{u}), & u \in D^+ \end{array} \right. \label{q2}\\
\varepsilon(u_1,u_2)& = &
\left \{ \begin{array}{ll} \displaystyle {1 \over 2} {\rm sgn} \, (u_1 - u_2), & u_1, u_2 \in \mathbb R \\
0, & {\rm otherwise} \end{array} \right.\label{q3}
\end{eqnarray}
and
\begin{eqnarray}
S(x,y) & = &  \sum_{j=0}^{M/2 - 1} {1 \over r_j} \Big ( q_{2j}(x) \tau_{2j+1}(y) -
q_{2j+1}(x) \tau_{2j}(y) \Big ) \label{S1} \\
D(x,y) & = &  \sum_{j=0}^{M/2 - 1} {1 \over r_j} \Big ( q_{2j}(x) q_{2j+1}(y) -
q_{2j+1}(x) q_{2j}(y) \Big ) \label{S2} \\
\tilde{I}(x,y) & = &  \sum_{j=0}^{M/2 - 1} {1 \over r_j} \Big ( \tau_{2j}(x) \tau_{2j+1}(y) -
\tau_{2j+1}(x) \tau_{2j}(y) \Big ) + \varepsilon(x,y). \label{S3} \\
\end{eqnarray}

According to working in \cite{BS07,Ma10}, the $(k_1 + k_2)$-point correlation function is given in terms of a Pfaffian by
\begin{eqnarray}\label{pf}
&&\rho_{(k_1+k_2)}(x_1,\dots,x_{k_1}; z_1,\dots,z_{k_2}) \nonumber \\
&& \qquad =
{\rm Pf} \begin{bmatrix} [K_{\rm rr}(x_j,x_l)]_{j,l=1,\dots,k_1} &
[K_{\rm rc}(x_j,z_l)]_{j=1,\dots,k_1 \atop l=1,\dots,k_2}  \\
[K_{\rm cr}(z_j,x_l)]_{j=1,\dots,k_1 \atop l=1,\dots,k_2} &
[K_{\rm cc}(z_j,z_l)]_{j=1,\dots,k_2 \atop l=1,\dots,k_2}
\end{bmatrix},
\end{eqnarray}
where
\begin{align*}
K_{\rm rr}(x,y) = 
\left [ \begin{array}{cc} S_{\rm rr}(x,y) & - D_{\rm rr}(x,y) \\
\tilde{I}_{\rm rr}(x,y) & S_{\rm rr}(y,x) \end{array} \right ] & \qquad
K_{\rm rc}(x,z) =
\left [ \begin{array}{cc} S_{\rm rc}(x,z) & - D_{\rm rc}(x,z) \\
\tilde{I}_{\rm rc}(x,z) & S_{\rm cr}(z,x) \end{array} \right ] \\
K_{\rm cr}(z,x) =
\left [ \begin{array}{cr} S_{\rm cr}(z,x) & - D_{\rm cr}(z,x) \\
\tilde{I}_{\rm cr}(z,x) & S_{\rm rc}(x,z) \end{array} \right ] & \qquad
K_{\rm cc}(z_1,z_2) =
\left [ \begin{array}{cc} S_{\rm cc}(z_1,z_2) & - D_{\rm cc}(z_1,z_2) \\
\tilde{I}_{\rm zz}(z_1,z_2) & S_{\rm cc}(z_2,z_1) \end{array} \right ]
\end{align*}
Here $S_{\cdot,\cdot}$, $D_{\cdot,\cdot}$, $\tilde{I}_{\cdot,\cdot}$ refer to (\ref{S1})--(\ref{S3});
the subscripts indicate if the corresponding variables are real or complex, which according to 
(\ref{q2}) and (\ref{q3}) effects the functional form

For future reference, we remark that (\ref{pf}) gives for the one and two point correlations of the real eigenvalues
\begin{eqnarray}
\rho_{(1)}^{\rm r}(x) & = & S_{\rm rr}(x,x) \nonumber \\
\rho_{(2)}^{\rm r}(x_1,x_2) & = & S_{\rm rr}(x_1,x_1) S_{\rm rr}(x_2,x_2) \nonumber \\
&& - \Big ( S_{\rm rr}(x_1,x_2) S_{\rm rr}(x_2,x_1) + D_{\rm rr}(x_1,x_2)
\tilde{I}_{\rm rr}(x_1,x_2) \Big ). \label{f2}
\end{eqnarray}
Also, structurally the same formulas hold for the one and two point correlations of the complex eigenvalues --- replace the subscripts rr by cc.

\subsection{Skew orthogonal polynomials and summation formulas}
The practical use of (\ref{pf}) requires the explicit form of the skew orthogonal polynomials, as determined by (\ref{g1g}) and (\ref{sk}). To proceed directly in their computation appears to be a formidable task. However, as observed in \cite{KSZ10}, and in \cite{Som07,SW08,APS08} for other ensembles of non-Hermitian random matrices, an indirect determination is possible.  This follows by evaluating the LHS of the identity
\begin{eqnarray}\label{V1}
&& (x - y) \langle \det (x - V_{(M-2) \times (M-2)}) (y - V_{(M-2) \times (M-2)}) \rangle \nonumber \\
&& \qquad = \sum_{j=0}^{M/2 - 1} {1 \over r_j} \Big (
p_{2j+1}(x) p_{2j}(y) - p_{2j+1}(y) p_{2j}(x) \Big ),
\end{eqnarray}
valid for $M$ even. Here $V_{(M-2)\times (M-2)}$ is the bottom right block in (\ref{UA}) in the case that the real orthogonal matrix $U$ is of size $(L+M-2) \times (L+M-2)$, and the average is over
$V_{(M-2)\times (M-2)}$. The result of \cite[eq.~(8)]{KSZ10} is that the LHS is equal to
\begin{equation}\label{V2}
(x - y) \sum_{m=0}^{M-2} {(L+m)! \over L! m!} (xy)^m.
\end{equation}

Substituting (\ref{V2}) in (\ref{V1}), replacing $M$ by $M+2$, then subtracting the original equation gives
\begin{eqnarray*}
&&{1 \over r_{M/2} }\Big ( p_{M+1}(x) p_M(y) - p_{M+1}(y) p_M(x) \Big ) \nonumber \\
&& \qquad = (x-y) \Big (
{(L+M)! \over L! M!} (xy)^M + {(L+M-1)! \over L! (M-1)!} (xy)^{M-1} \Big ).
\end{eqnarray*}
Recalling that each $p_j(x)$ is monic of degree $j$, and that $M$ is even, it follows from this that
\begin{equation}\label{V3}
p_j(x) = \left \{ \begin{array}{ll} x^j ,& j \quad {\rm even} \\
\displaystyle x^j - {j-1 \over L + j - 1} x^{j-2}, & j \quad {\rm odd} \end{array} \right.
\end{equation}
and
\begin{equation}
{1 \over r_j} = {(L+2j)! \over L! (2j)!}.
\end{equation}
Furthermore, comparing the RHS of (\ref{V1}) with the definition (\ref{S2}) of $D(x,y)$, it follows from (\ref{S2}) that
\begin{equation}
D(x,y) = - \tilde{w}(x) \tilde{w}(y) (x - y) \sum_{m=0}^{M-2} {(L+m)! \over L! m!} (xy)^m.
\end{equation}
Recognizing the sum as a partial sum of the generalized binomial series, we obtain the summation fomula
\begin{equation}\label{Dxy}
\lim_{M \to \infty} D(x,y) = - {\tilde{w}(x) \tilde{w}(y) (x-y) \over (1 - xy)^{L+1}}.
\end{equation}

From knowledge of (\ref{V3}) we would like to similarly obtain explicit forms for the quantities $S$ and $\tilde{I}$ as specified by (\ref{S1}) and (\ref{S3}) respectively. When both variables are complex, this is immediate, as we can check from the definitions that
\begin{equation}\label{S2}
S_{\rm cc}(z_1,z_2) = i D(z_1,\bar{z}_2), \qquad
\tilde{I}_{\rm cc}(z_1,z_2) = - D(\bar{z}_1,\bar{z}_2).
\end{equation}
Applying (\ref{Dxy}) we thus have
\begin{eqnarray}\label{S3}
\lim_{M \to \infty} S_{\rm cc}(z_1,z_2) & = & - i {\tilde{w}(z_1) \tilde{w}(z_2) (z_1 - \bar{z}_2) \over
(1 - z_1 \bar{z}_2)^{L+1}} \nonumber \\
\lim_{M \to \infty} \tilde{I}_{\rm cc}(z_1,z_2) & = &  {\tilde{w}(z_1) \tilde{w}(z_2) (\bar{z}_1 - \bar{z}_2) \over
(1 - \bar{z}_1 \bar{z}_2)^{L+1}}.
\end{eqnarray}

Consider next the case that both variables are real. Our strategy to compute $S_{\rm rr}(x,y)$ is borrowed from \cite{FN08p}. There structures we are about to exhibit were observed in the case of the partially symmetric real Ginibre ensemble. Thus we begin by observing from (\ref{V3}) that
\begin{equation}\label{pn}
p_{2n+1}(x) = - {1 \over L + 2n} (1 - x^2)^{-L/2 +1 } {d \over dx}
\Big ( (1 - x^2)^{L/2} x^{2n} \Big ).
\end{equation}
Recalling the definition (\ref{q2}) then shows
$$
\tau_{2n+1}(x) = {1 \over L + 2n} \Big ( {L \over 2 \pi}
{\Gamma(1/2) \Gamma((L+1)/2) \over \Gamma(L/2)} \Big )^{1/2} (1 - x^2)^{L/2} x^{2n},
$$
and so
\begin{eqnarray}\label{C1}
\sum_{j=0}^{M/2-1} {1 \over r_j} q_{2j}(x) \tau_{2j+1}(y) & = &
{L \over 2 \pi} {\Gamma(1/2) \Gamma((L+1)/2) \over \Gamma(L/2)}(1 - x^2)^{L/2-1} \nonumber \\
&& \times
(1 - y^2)^{L/2} \sum_{j=0}^{M/2 - 1} {1 \over r_j} {1 \over L + 2j} p_{2j}(x) p_{2j}(y).
\end{eqnarray}
Also
\begin{eqnarray}\label{tt1}
\lefteqn{\sum_{j=0}^{M/2} {1 \over r_j} q_{2j+1}(x) \tau_{2j}(y)} \nonumber \\
 & & =
\sum_{j=0}^{M/2 - 1} {1 \over r_j} \tau_{2j}(y) \Big ( x^{2j+1} - {2j \over L + 2j}
x^{2j-1} \Big ) 
 \nonumber
\\
&&
= {1 \over r_{M/2 - 1}} \tau_{M-2}(y) \tilde{w}(x) x^{M-1}  \nonumber
\\
&& \quad - \tilde{w}(x)
\sum_{j=0}^{M/2 - 2}
{(L+2j+1)! \over L! (2j+1)!} x^{2j+1}
\Big ( \tau_{2j+2}(y) - {2j+1 \over L + 2j + 1} \tau_{2j}(y) \Big ).
\end{eqnarray}

To proceed further, we note that analogous to (\ref{pn}) we have 
\begin{eqnarray*}
\lefteqn{p_{2j+2}(y) - {2j + 1 \over L + 2j + 1} p_{2j}(y) }\\
&& \qquad = - {1 \over L + 2j + 1} (1 - y^2)^{-L/2 + 1} {d \over dy}
\Big ( (1 - y^2)^{L/2} y^{2j+1} \Big ).
\end{eqnarray*}
Recalling now the definition (\ref{q2}) of $\tau_k$ we see
\begin{equation}\label{tt2}
\tau_{2j+2}(y) - {2j + 1 \over L + 2j + 1} \tau_{2j}(y) =
{1 \over L + 2j + 1} \tilde{w}(y) y^{2j+1} (1 - y^2).
\end{equation}
Substituting (\ref{tt2}) in (\ref{tt1}) shows
\begin{eqnarray}\label{C2}
&&\sum_{j=0}^{M/2 - 1} {1 \over r_j} p_{2j+1}(x) \tau_{2j}(y) =
{1 \over r_{M/2-1}} \tau_{M_2}(y) \tilde{w}(x) x^{M-1} \nonumber \\
&& \qquad -
\tilde{w}(x) \tilde{w}(y) (1 - y^2) \sum_{j=0}^{M/2 - 2}
{(L+2j+1)! \over L! (2j+1)!} {1 \over L + 2j+1}  x^{2j+1}.
\end{eqnarray}
Appropriately combining (\ref{C1}) and (\ref{C2}) then gives our sought summation formula
\begin{eqnarray*}
&& \lefteqn{S_{\rm rr}(x,y) = - {1 \over r_{M/2 - 1}} \tau_{M-2}(y)\tilde{w}(x) x^{M-1} }\nonumber \\
&& \qquad + {1 \over \pi}
{\Gamma(1/2) \Gamma((L+1)/2 \over \Gamma(L/2)} (1 - x^2)^{L/2-1}
(1 - y^2)^{L/2}
\sum_{j=0}^{M-2} {(L+j-1)! \over (L-1)! j!} x^j y^j.
\end{eqnarray*}
Finally, taking the $M \to \infty$ limit and relating the resulting sum to the generalized binomial series we obtain
\begin{equation}\label{Srr}
\lim_{M \to \infty} S_{\rm rr}(x,y) =
{1 \over \pi} {\Gamma(1/2) \Gamma((L+1)/2) \over \Gamma(L/2)}
{(1 - x^2)^{L/2 - 1} (1 - y^2)^{L/2} \over (1 - xy)^L}.
\end{equation}

With $S_{\rm rr}(x,y)$ known, $\tilde{I}_{\rm rr}(x,y)$ follows from the formula
\begin{equation}\label{Sp}
\tilde{I}_{\rm rr}(x,y) = - \int_x^y S_{\rm rr}(u,y) \, du +
{1 \over 2} {\rm sgn} \, (x-y),
\end{equation}
which in turn follows from the definitions (\ref{S3}), (\ref{S1}) and (\ref{q2}).
We note that in the case $L=1$ the integral in (\ref{Sp}) with the substitution
(\ref{Srr}) can be evaluated to give
\begin{equation}\label{Is}
\lim_{M \to \infty} \tilde{I}_{\rm rr}(x,y) \Big |_{L=1} = {{\rm sgn} \, (y-x) \over \pi}
\arcsin {\sqrt{(1-x^2)(1-y^2)} \over 1 - xy}.
\end{equation}

Calculation of the quantities $S_{\cdot,\cdot}$ and $\tilde{I}_{\cdot,\cdot}$ for the
real/complex and complex/real blocks requires no further work. Thus it follows from the definitions that
\begin{eqnarray}\label{Sp1}
S_{\rm cr}(z,x) & = & S_{\rm rr}(z,x) \nonumber \\
S_{\rm rc}(x,z) & = & i D_{\rm rc}(x,\bar{z}) \nonumber \\
\tilde{I}_{\rm cr}(z,x) & = & - \tilde{I}_{\rm rc}(x,z) \: = \: i S_{\rm rr}(\bar{z},x).
\end{eqnarray}

\section{Conclusion}
From Section 2.1 we know that the zeros of the matrix-valued real Gaussian power series (\ref{1}) are statistically identical to the eigenvalues of the truncated real orthogonal matrices $V_{M \times M}$, $M \to \infty$ in (\ref{UA}).
The correlations for the latter are given by the Pfaffian (\ref{pf}), with entries specified by (\ref{Dxy}), (\ref{S2}),
(\ref{S3}), (\ref{Srr}) and (\ref{Sp}).

In the case $L=1$ the matrix valued random function in (\ref{1}) is equal to the $N \to \infty$ limiting form of the Kac random polynomial (\ref{0}). The explicit form of the entries in (\ref{pf}) are then
\begin{eqnarray}
 \lefteqn{K_{\rm rr}(x,y) \Big |_{L=1}} \nonumber \\
&& \quad = {1 \over \pi}
\begin{bmatrix} \displaystyle {(1-y^2)^{1/2} \over (1 - x^2)^{1/2} (1 - xy)} & \displaystyle {(x-y) \over (1 - x^2)^{1/2} (1 - y^2)^{1/2} (1 - xy)^2} \\ \displaystyle
 {\rm sgn}(y-x) \arcsin {\sqrt{(1-x^2)(1-y^2)} \over 1 - xy} & \displaystyle {(1-x^2)^{1/2} \over (1 - y^2)^{1/2} (1 - xy)}
\end{bmatrix}, \nonumber \\
\end{eqnarray}
\begin{eqnarray}
 \lefteqn{K_{\rm rc}(x,z) \Big |_{L=1}} \nonumber \\
&& \quad = {1 \over \pi}
\begin{bmatrix} \displaystyle  {-i (x-\bar{z}) \over (1 - x^2)^{1/2} |1 - z^2|^{1/2} (1 - x\bar{z})^2} & 
\displaystyle {(x-z) \over (1 - x^2)^{1/2} |1 - z^2|^{1/2} (1 - xz)^2} \\ \displaystyle
  {-i (1-x^2)^{1/2} \over |1 - \bar{z}^2|^{1/2} (1 - \bar{z} x) } & \displaystyle {(1-x^2)^{1/2} \over |1 - {z}^2|^{1/2} (1 - {z} x) }
\end{bmatrix}, \nonumber \\
\end{eqnarray}
\begin{eqnarray}
 \lefteqn{K_{\rm cr}(z,x) \Big |_{L=1}} \nonumber \\
&& \quad = {1 \over \pi}
\begin{bmatrix} \displaystyle {(1-x^2)^{1/2} \over |1 - {z}^2|^{1/2} (1 - {z} x) }
 & 
\displaystyle {-(x-z) \over (1 - x^2)^{1/2} |1 - z^2|^{1/2} (1 - xz)^2} \\ \displaystyle
  {i (1-x^2)^{1/2} \over |1 - \bar{z}^2|^{1/2} (1 - \bar{z} x) } & \displaystyle  {-i (x-\bar{z}) \over (1 - x^2)^{1/2} |1 - z^2|^{1/2} (1 - x\bar{z})^2}
\end{bmatrix}, \nonumber \\
\end{eqnarray}
\begin{eqnarray}
 \lefteqn{K_{\rm cc}(z_1,z_2) \Big |_{L=1}} \nonumber \\
&& \quad = {1 \over \pi}
\begin{bmatrix} \displaystyle {-i(z_1 - \bar{z}_2) \over |1 - z_1^2|^{1/2} |1 - {z}_2^2|^{1/2} (1 - {z}_1 \bar{z}_2)^2 }
 & 
\displaystyle {(z_1-z_2) \over |1 - z_1^2|^{1/2} |1 - z^2|^{1/2} (1 - z_1z_2)^2} \\ \displaystyle
\displaystyle {(\bar{z}_1-\bar{z}_2) \over |1 - z_1^2|^{1/2} |1 - z_2^2|^{1/2} (1 - \bar{z}_1\bar{z}_2)^2}   & \displaystyle  {i 
(\bar{z}_1 - {z}_2) \over |1 - z_1^2|^{1/2} |1 - {z}_2^2|^{1/2} (1 - \bar{z}_1 z_2)^2 }
\end{bmatrix}, \nonumber \\
\end{eqnarray}

Recalling (\ref{f2}), we check from these explicit forms that the results (\ref{4}), (\ref{4a}) for the one and two
point real correlations and the result (\ref{zc}) for the one point complex correlation are reclaimed. These results
were obtained from the forms (\ref{rr}) and (\ref{Hf}), different from the corresponding
Pfaffian formula (\ref{pf}). An outstanding task then is to give a direct derivation of the equivalence between the
forms. In the case of complex limiting Kac polynomials, the analogous problem was solved by
Peres and Vir\'ag \cite{PV03}.

\section*{Acknowledgements}
This work was done while the author was a member of the MSRI, participating in the Fall 2010 program `Random matrices, interacting particle systems and integrable systems'. Partial support for this reseach was also provided by the Australian Research Council.


\input{KacPoly.bbl}

\end{document}

%% file: KacPoly.bbl
\providecommand{\bysame}{\leavevmode\hbox to3em{\hrulefill}\thinspace}
\providecommand{\MR}{\relax\ifhmode\unskip\space\fi MR }
\providecommand{\MRhref}[2]{%
  \href{http://www.ams.org/mathscinet-getitem?mr=#1}{#2}
}
\providecommand{\href}[2]{#2}